\documentclass[prl,twocolumn,floatfix,amsmath,amssymb]{revtex4}
\pdfoutput=1
\usepackage{graphicx}
\usepackage{dcolumn}
\usepackage{bm}
\usepackage{amssymb}
\usepackage{amsmath}
\usepackage{color}
\renewcommand{\vec}{\mathbf}

\begin{document}

\title{Dissipation and criticality in the  lowest Landau level of graphene}

\author{Xun Jia}
\author{Pallab Goswami}
\author{Sudip Chakravarty}
\affiliation{Department of Physics and Astronomy, University of
California Los Angeles, Los Angeles, CA 90095-1547}

\date{\today}

\begin{abstract}
The lowest Landau level of graphene is studied numerically
by considering a tight-binding Hamiltonian with disorder. The Hall
conductance $\sigma_\mathrm{xy}$ and the longitudinal conductance
$\sigma_\mathrm{xx}$ are computed. We demonstrate that bond disorder can produce a  plateau-like feature centered at 
$\nu=0$, while the longitudinal conductance is  nonzero
in the same  region, reflecting a band of extended states between $\pm E_{c}$, whose magnitude depends on the disorder strength. The  critical exponent  corresponding to the  localization length at the edges of this band is found to be $2.47\pm 0.04$. When both bond disorder and a finite mass term exist the localization length exponent varies continuously  between $\sim 1.0$ and $\sim 7/3$.

\end{abstract}


\maketitle 

The quantum Hall effect (QHE), either integer or
fractional, is one of the most intriguing phenomena in physics~\cite{Laughlin:1999} that
has drawn, and still continues to draw, enormous attention even after two decades since its discovery. In graphene unconventional
QHE~\cite{Novoselov:2005,Zhang:2005:NAT} corresponding to
 $\sigma_\mathrm{xy}=\pm(n+1/2)4 (e^2/h)= \pm 4\nu (e^2/h),\; n=0,1, 2, \ldots$, where $\sigma_{xy}$ is the Hall conductance,
has been addressed in terms of low lying excitations akin to
relativistic Dirac fermions~\cite{Gusynin:2005,Peres:2006}; the
factor of 4 arises from the two-fold valley and spin degeneracies; $e$ is the
electronic charge and $h$ the Planck's constant. Yet, a more complete theory of
QHE requires an understanding of the localization-delocalization
transitions within the Landau levels, which reflects a very special
quantum phase transition, especially in graphene with its low energy
Dirac spectra~\cite{Ostrovsky:2007,Goswami:2007,Koshino:2007}. It is
indeed paradoxical that such a precise quantization requires
material defects and disorder, and even the nature of disorder seems
to matter for  graphene.

A remarkable recent discovery in graphene is, what appears to be, a  $\nu=0$ plateau in
a sufficiently large magnetic field~\cite{Zhang:2006,Abanin:2007}, although the plateau in the measured $\rho_{xy}$ is difficult to decipher.  The
longitudinal resistivity $\rho_{\mathrm{xx}}$, on the other hand, is observed to be
nonzero ($\gtrsim h/e^2$) in the region of the plateau~\cite{Abanin:2007}, in
sharp contrast to the non-dissipative behavior of
$\rho_{\mathrm{xx}}$ in conventional QHE, crying out for a
theoretical explanation. It  is most peculiar because ``a
dissipative quantum Hall plateau'' is an oxymoron, for, according to
Laughlin ~\cite{Laughlin:1999}, the precise quantization requires
zero dissipation.  An intriguing explanation of this paradoxical
phenomenon is given in Ref.~\cite{Abanin:2007}, where the removal of
spin degeneracy plays a special role, and the nonzero longitudinal
resistivity $\rho_\mathrm{xx}$ is ascribed to a pair of gapless
counter-propagating chiral edge modes carrying opposite spins, while a spin gap in the 
bulk protects the quantization of the Hall conductance. The role of interactions is crucial 
in this theory.

Surprisingly, in the present  Letter we shall
demonstrate by explicitly computing $\sigma_{xy}$, and the
longitudinal conductance $\sigma_{xx}$, that the existence of a plateau-like feature at $\nu=0$
and the non-zero $\rho_{xx}$ can be attributed
to a single mechanism, an inter-valley coupling that is induced by
bond disorder. Because
$\sigma_{xy}$ is  zero at $\nu=0$,
$\rho_{xx}=1/\sigma_{xx}$ at the same point. The most remarkable feature is that there is a band of extended states centered at zero energy, whose extent depends on the strength of bond disorder. In contrast to  Ref.~\cite{Abanin:2007}, in our picture there is no quantization of $\sigma_{xy}$ at $\nu=0$, only a sloping behavior as a function of energy. In addition, dissipation is not an edge phenomenon, but a bulk one. 

Another remarkable aspect of the integer QHE in graphene is the
possible existence of a continuously varying critical exponent of
the divergence of the localization length within the Landau band
for a class of disorder. We show: (1) If there is only bond
disorder, the critical exponent is close to $\sim 7/3$, same as in
the conventional integer QHE; (2) when a finite mass term is added
in addition to bond disorder, the critical exponent  depends on the
ratio of the intensity of bond disorder
 to the finite mass, continuously varying from $\sim
1.0$ to $\sim 7/3$, as the system is tuned from weak to strong
disorder, which correctly reaffirms the results of Ref.~\cite{Goswami:2007}
obtained by an entirely different method; the present method is
more powerful, because larger system sizes can be handled.  A finite mass appears to be experimentally relevant.
Although the removal of the nodal degeneracy in the lowest
Landau level may be of  many body origin, it can be approximately
accounted for by including an explicit mass term in the Hamiltonian.

We study the tight-binding model of graphene subject
to a constant perpendicular magnetic field and disorder~\cite{Sheng:2006}, using
real space transfer matrix and exact
diagonalization methods, to explore  the peculiar properties of
the lowest Landau level discussed above.
\begin{figure}
    \centering
  \includegraphics[width=7.5cm]{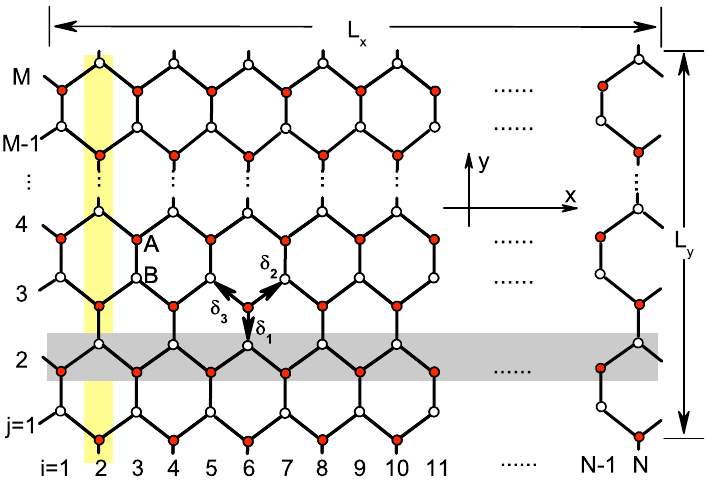}\\
  \caption{(Color online) Graphene lattice. Lattice sites belonging to the
  same shaded horizontal or vertical stripes share the same indices $i$ or $j$.
  The solid circles correspond to the $A$ sublattice and the open circles to the
  $B$ sublattice. The three nearest neighbor vectors joining the two sublattices
  are $\boldsymbol{\delta}_{k}$, $k=1 \ldots 3$.}
  \label{lattice}
\end{figure}
The Hamiltonian defined on a honeycomb lattice  of
dimension $L_x\times L_y$, shown in Fig.~\ref{lattice},  is
\begin{equation}\label{hamtonian}
\begin{split}
    H= & \sum_{\vec{n}} [(\epsilon_{\vec{n},A}+m) c_{\vec{n},A}^\dag c_{\vec{n},A} +
    (\epsilon_{\vec{n},B} -m)c_{\vec{n},B}^\dag c_{\vec{n},B}]\\
    &- \sum_{\vec{n}}\sum_{k=1}^{3}
    (t_{\vec{n},k}\mathrm{e}^{\mathrm{i} a_{\vec{n},k}}c_{\vec{n},A}^\dag
    c_{\vec{n+\boldsymbol{\delta}}_k,B}+\mathrm{h.c.}),
\end{split}
\end{equation}
where the summation of $\vec{n}$ ranges over all unit cells, and
$c_{\vec{n},A}$, $c_{\vec{n},B}$ are the fermionic annihilation
operators in the unit cell $\vec{n}$ for the sublattices $A$ and
$B$, respectively.  The spacing between vertical slices is $\sqrt{3}a/2$, where $a$ is the bond length. In the transfer matrix calculations $a$ is the unit of length. The size of the sample is chosen such that $L_x=N(\sqrt{3}a/2)$ and $L_y=M(3a)$, where $N$ is the number of vertical slices and $M$ is the number of unit cells on a vertical slice.

The spin degrees of freedom are omitted, as we
assume that the magnetic field is sufficiently strong to completely
polarize them. The on-site energies $\epsilon_{\vec{n},A}$ and
$\epsilon_{\vec{n},B}$ are independent random variables. Thus, $V_{\vec{n}}=(\epsilon_{\vec{n},A}+\epsilon_{\vec{n},B})/2$ is a random potential and  $M_{\vec{n}}=(\epsilon_{\vec{n},A}-\epsilon_{\vec{n},B})/2$ is mass disorder in the corresponding language of  the low energy spectra of Dirac fermions. Here we do not consider  mass disorder and therefore choose $\epsilon_{\vec{n},A}=\epsilon_{\vec{n},B}= V_{\vec{n}}$, where $V_{\vec{n}}$ is uniformly distributed in the range $[-g_V/2, g_V/2]$. The mass $m$ provides a charge density modulation between the two sublattices and leads to an energy gap. In the Dirac  language  this  gap would appear as a parity-preserving mass. We choose $t_{\vec{n},k}=t+\Delta t_{\vec{n},k}$ and set  $t=1$, providing a
natural energy scale. The quantity $\Delta t_{\vec{n},k}$ is a
random variable uniformly distributed in the range $[-g_T/2,g_T/2]$, characterizing the bond disorder. The phases
$a_{\vec{n},k}$  are such that the magnetic flux per
hexagonal plaquette, $\phi$, is $1/Q$, in units of the flux quanta
$\phi_0=h/e$. We choose a gauge such that $a_{\vec{n},1}=\pi i/Q$
for the vertical bonds in slice $i$ as in Fig.~\ref{lattice}, and
$a_{\vec{n},2}=a_{\vec{n},3}=0$.

The longitudinal conductance $\sigma_\mathrm{xx}$ is studied using
 the well developed transfer matrix method. Consider a
quasi-1D system, $L_x\gg L_y$ with a periodic boundary
condition only along the $y$ direction. Let $\Psi_i =
(\psi_{i,1},\psi_{i,2}, \ldots, \psi_{i,2M})^T$ be the amplitudes
on the slice $i$ for an eigenstate with a given energy $E$; then amplitudes on the successive slices are related
by the matrix multiplication:
\begin{equation}
    \label{transfermatrix}
    \left[
      \begin{array}{c}
        \Psi_{i+1} \\
        \Psi_{i} \\
      \end{array}
    \right] = \left[
                \begin{array}{cc}
                  {\cal T}^{-1}_i(E-H_i) & -{\cal T}_i^{-1}{\cal T}_{i-1} \\
                  1 & 0 \\
                \end{array}
              \right]\left[
                       \begin{array}{c}
                         \Psi_i \\
                         \Psi_{i-1} \\
                       \end{array}
                     \right],
\end{equation}
where ${\cal T}_i$ is a diagonal matrix with elements
$(t_{i,1},t_{i,2},\ldots,t_{i,2M})$ representing the hopping matrix
elements connecting the slices $i$ and $i+1$, and $H_i$ is the
Hamiltonian within the slice. All postive Lyapunov exponents of the
transfer matrix~\cite{Kramer:1996},
$\gamma_1>\gamma_2>\ldots>\gamma_{2M}$, are computed by iterating
Eq.~(\ref{transfermatrix}) and performing frequent orthonormalizations. The convergence of this algorithm is guaranteed by the
well known Osledec theorem~\cite{Oseledec:1968}. The
conductance per square, $\sigma_\mathrm{xx}$, is given by the Landauer
formula\cite{Fisher:1981,Baranger:1989,Kramer:1993,Sheng:2000}(note the special factor of $\sqrt{3}$ in the argument of $\cosh$):
\begin{equation}\label{landauer}
    \sigma_\mathrm{xx}=\frac{e^2}{h}\sum_{i=1}^{2M} \frac{1}{\cosh^2
    (2\sqrt{3}M\gamma_i)}.
\end{equation}
The localization length in the
quasi-1D system of width $L_y$ is given by
$\lambda_M = 1/\gamma_{2M}$. Assuming single parameter scaling,  $\lambda_M/M = f(|
E-E_c|M^{1/\nu_{\ell}})$, the data collapse yields the
critical exponent $\nu_{\ell}$ and the critical energy $E_c$.

To compute the Hall conductance $\sigma_\mathrm{xy}$, we impose
periodic boundary conditions in both directions of the system. The
Hamiltonian (\ref{hamtonian}) is diagonalized to obtain a set of
energy eigenvalues $E_\alpha$ and the corresponding set of
eigenstates $|\alpha\rangle$ for $\alpha=1,\ldots,2M\times N$.
Then $\sigma_\mathrm{xy}$ is computed using the Kubo formula
\cite{Thouless:1982}:
\begin{equation}\label{kubo}
    \sigma_\mathrm{xy}(E)=\frac{ie^2\hbar}{L_xL_y}\sum_{E_\alpha < E < E_\beta}
    \frac{\langle \alpha|v_y|\beta\rangle\langle\beta|v_x|\alpha\rangle-
  ( x\leftrightarrow y)}{(E_\alpha-E_\beta)^2},
\end{equation}
where $v_x=[H,x]/i\hbar$ is the velocity operator along the $x$
direction and  similarly for $v_y$ in the $y$ direction. Note that the bonds in
Fig.~\ref{lattice} that are not parallel to the $y$ direction contribute
to both $v_x$ and $v_y$. The summation corresponds to sum over the states
below and above the energy $E$. Finally, the
expression is disorder averaged.

\begin{figure}[htb]
    \centering
  \includegraphics[width=8.5cm]{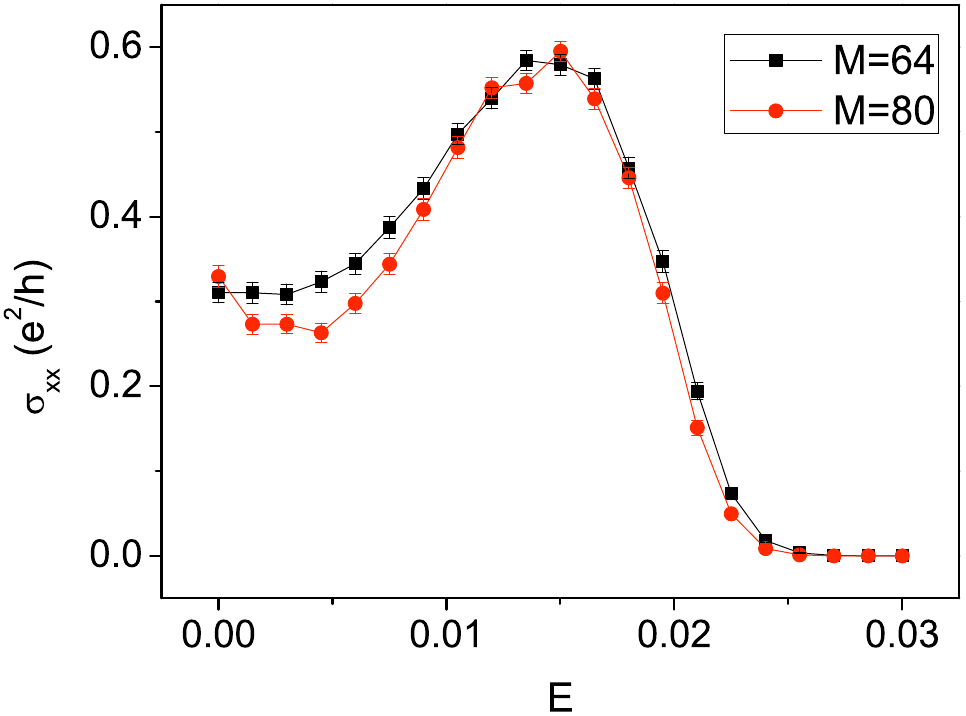}\\
  \caption{(Color online) The longitudinal conductance $\sigma_\mathrm{xx}$  as a function of
  energy $E$. The magnetic flux through the hexagonal plaquette  $\phi=1/200$
 and  $g_T=0.5$.}
  \label{Fig2}
\end{figure}

\begin{figure}
  \includegraphics[width=8cm]{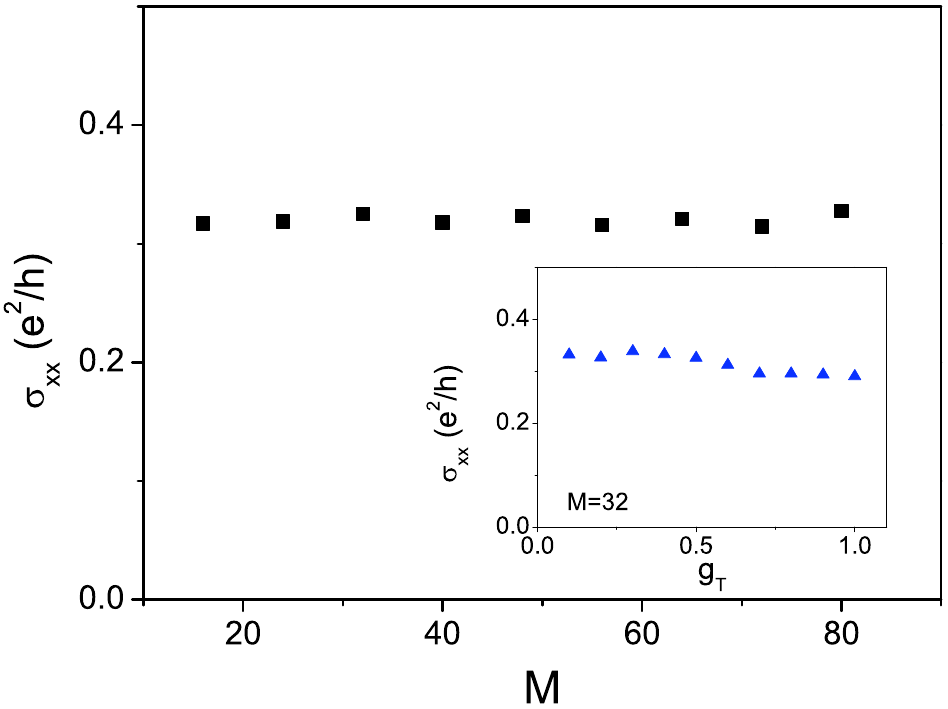}\\
  \caption{(Color online) Longitudinal conductance $\sigma_\mathrm{xx}$ at $E=0$ as
  a function of the transverse size $M$ for $g_T=0.5$ and $g_V=0$. The inset shows $\sigma_{\mathrm{xx}}(E=0)$ as a function of $g_{T}$ for $M=32$. The value appears to be independent of $g_{T}$ and close to $\frac{e^{2}}{\pi h}$; the errors bars are about the size of the scatter in the numerical data. The near universality disappears if $g_{V}$ is added.}
  \label{sigma_xx}
\end{figure}

In the language of Dirac fermions~\cite{Goswami:2007} random hopping gives rise to both intranode and internode scattering between states on different sublattices. Intranode scattering appears as a random abelian gauge field, and the two inequivalent nodes have opposite charges corresponding to this gauge field. When projected to the lowest Landau level, the abelian gauge field leaves it unaffected. However, the internode scattering mixes the degenerate states corresponding to the two  inequivalent nodes and  produces extended states at $\pm E_c$. The existence of extended states at energies symmetric about $E=0$ is the consequence of the sublattice symmetry of the disorder (often referred to as the chiral or  the particle-hole symmetry). It is this special symmetry that leads to a divergent density of states and delocalized states at $E=0$ \cite{Goswami:2007,Hikami:1993}. However, the calculated  finite $\sigma_{xx}$ and a linear  variation of  $\sigma_{xy}$ with a small slope in the energy range $-E_c<E<E_c$ hint at the existence of a band of delocalized states between $\pm E_c$, as shown below.

 In the transfer matrix calculation of $\sigma_{\mathrm{xx}}$ (see Fig.~\ref{Fig2}), we chose
 a magnetic field $\phi=1/200$. An iteration
of Eq.~(\ref{transfermatrix}) of the order of $10^5$ to $10^6$ was
performed until the relative errors of less than $0.5\%$ of all the
Lyapunov exponents were achieved. The longitudinal conductance
$\sigma_\mathrm{xx}$ at exactly $E=0$ is computed according to the
Landauer formula, Eq.~(\ref{landauer}), in systems with different
values of $M$, as shown in Fig.~\ref{sigma_xx}. For only
bond disorder, $g_T=0.5$, a non-zero longitudinal conductance is
observed for all system sizes. The value of $\sigma_\mathrm{xx}\sim
\frac{e^{2}}{\pi h}$ is found to be independent of $M$ and $g_{T}$. This behavior of
$\sigma_{\mathrm{xx}}$ implies an unusual dissipative nature. We have also checked the
existence of the dissipative behavior when  $g_{v}\ne 0$
in addition. 
\begin{figure}[htb]
  \includegraphics[width=8cm]{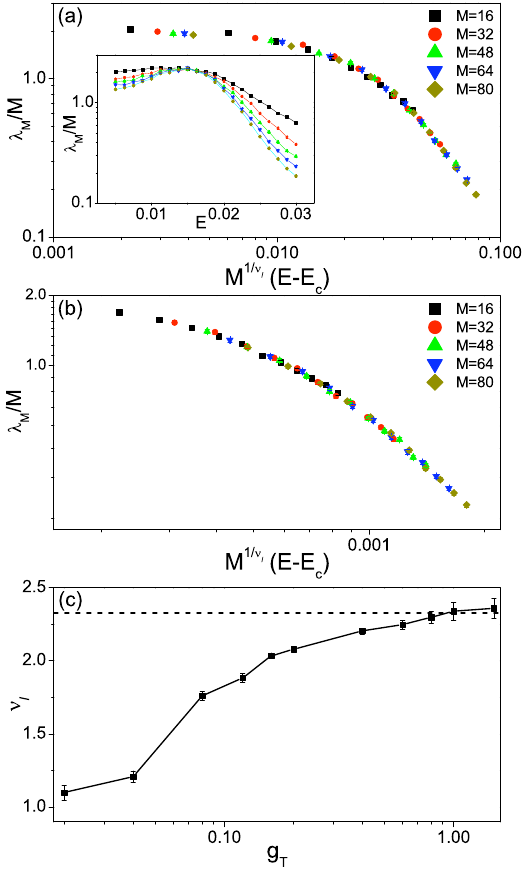}\\
  \caption{(Color online)(a) Scaling curve for the case $g_T=0.5$ and $m=0$.
  Insert shows the size dependence of the localization length computed in systems with
  different sizes. (b) Scaling curve for the case $g_T=0.2$ and $m=0.2$. (c)The
  critical exponent $\nu_{\ell}$ as a function of
  the random hopping intensity $g_T$ for $m=0.2$. It varies continuously
  from $\nu_{\ell}= 1.10\pm 0.05$ to $\nu= 2.36\pm 0.07$ as the disorder intensity
  $g_T$ is tuned from the weak to strong. The dashed line corresponds to  $\nu_{\ell}=7/3$.}
  \label{numgT}
\end{figure}

To the study of the critical
behavior in the presence of  bond disorder in the massless  case, we set $g_T=0.5$. The critical
exponent is expected to be independent of the value $g_T$~\cite{Goswami:2007}. The
renormalized localization lengths $\lambda_M/M$ as a function of $E$ for various $M$ are plotted in
in the insert of Fig.~\ref{numgT}(a). The critical energy $E_c$ is located at a
non-zero value $E_c=0.0167$ where $\lambda_M/M$ is
independent of $M$. A successful data collapse based on the data
with $E>E_c$ leads to a critical exponent of $\nu=2.47\pm 0.04$,
close to  conventional  integer QHE. The scaling form is
depicted in Fig.~\ref{numgT}(a).

For the massive case, we vary bond disorder with
a fixed $m=0.2$ for the purpose of illustration. An example of data collapse is shown
in Fig.~\ref{numgT}(b) with $g_T=0.2$.  The critical  exponent 
$\nu_{\ell}=2.08\pm 0.01$ is different from conventional QHE; $\nu_{\ell}$
varies continuously  from $1.1\pm 0.05$ to $2.36\pm 0.07$ as
the system is tuned from  $g_T=0.02$ to  $g_T=1.5$. This behavior
agrees with the previous results
~\cite{Goswami:2007}. When $g_T\gg m$,
the effect of the finite mass is negligible, and the critical
exponent of $\nu_{\ell}\approx 7/3$ is recovered, as before.

\begin{figure}[htb]
  \includegraphics[width=8cm]{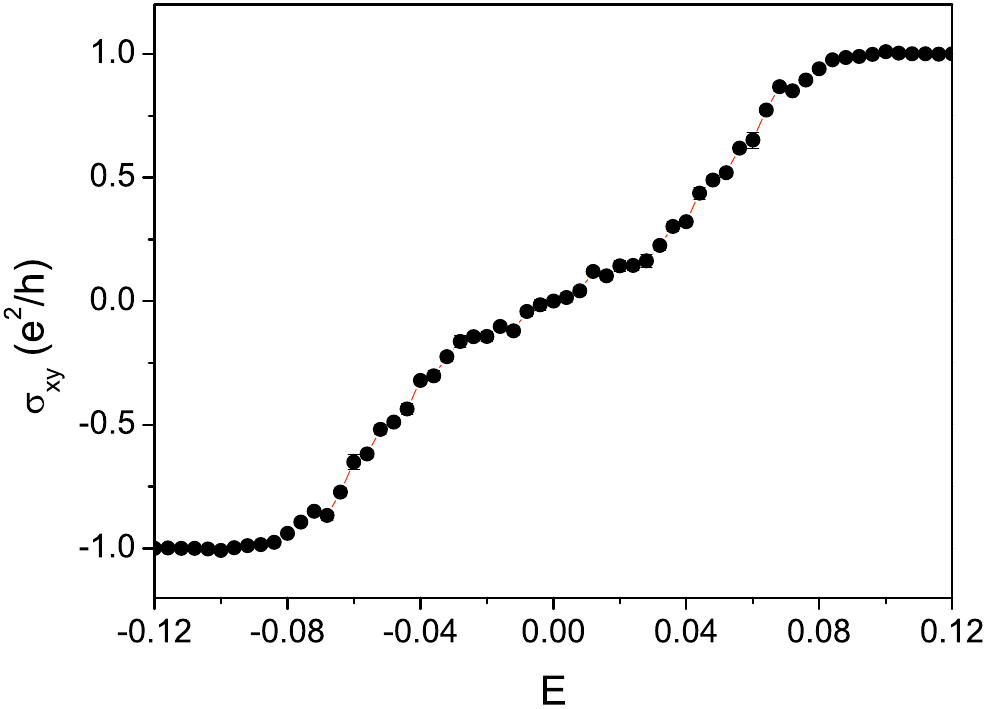}\\
  \caption{(Color online) Hall conductance $\sigma_\mathrm{xy}$ as a function
  of energy $E$ with bond disorder intensities $g_T$. The $\nu=0$
  plateau emerges due to the bond disorders.}
  \label{sigma_xy}
\end{figure}

In computing  the energy
dependence of $\sigma_\mathrm{xy}$ a relatively large flux $\phi=1/20$ is chosen, because
smaller values of flux involve many Landau bands in the diagonalization calculations, which
are hard to track accurately. The
chosen system size was $N=M=40$, and  an average over 1000 disorder
realizations was performed. The results are shown in Fig.~\ref{sigma_xy}.
Although there is not a strict plateau at $\nu =0$, there is a break in the slope
in the rise between  $\nu=-1$ to $\nu=1$, as the
energy sweeps past the band center at $E=0$. This can be construed as a plateau-like feature. Since the lowest Landau level splitting is expected
to increase with $g_T$, so is the extent of the region around $E=0$ with a smaller slope.

The striking results here are the band of extended states in the region $-E_{c}< E <E_{c}$ 
for bond disorder and  the vindication of a continuously varying localization length exponent
when, in addition, there is a finite uniform mass present in the Dirac spectrum. The situation 
is a bit more subtle, however.
We
had  previously observed that the density of states diverges very weakly
at $E=0$~\cite{Goswami:2007}. In  particular, for the Lorentzian
distribution of disorder this divergence was found to be exactly
logarithmic. A $\log^{2}E$ divergence
was predicted in Ref.~\cite{Hikami:1993}  for Gaussian disorder
corresponding to a  Hamiltonian which in fact is formally identical
when projected to the lowest Landau level. Such a weak divergence at $E=0$ may
give rise to fluctuations responsible for dissipation leading to a
finite $\sigma_{\mathrm{xx}}$~\cite{Hikami:1993,Ostrovsky:2007}. 
The  slightly sloping profile of $\sigma_{\mathrm{xy}}$
centered at $E=0$  is harder to explain analytically. The dissipative behavior 
at $\nu=0$ is consistent with experiments. However, in contrast to Ref.~\cite{Abanin:2007}, this
dissipative behavior is a bulk phenomenon, not an edge phenomenon. It is possible to test this 
experimentally by varying the aspect ratio of the sample. But we only have a sloping plateau-like feature
of $\sigma_{xy}$ at $\nu=0$ unlike Ref.~\cite{Abanin:2007}.   It is possible 
that the difference between the two pictures depends on the relative size of the spin splitting
compared to the width of the extended band of states. Clearly further experimental and theoretical
work would be very helpful  to elucidate the precise nature of this exciting new development.

This work is supported by NSF under Grant No. DMR-0705092. We thank P. A. Lee for valuable comments.

\end{document}